\documentclass{JHEP3}

  \usepackage{latexsym,bm,amsmath,amssymb,amsfonts}
  \usepackage{epsfig,graphics,graphicx}
  \usepackage{slashed}
  \usepackage[latin1]{inputenc}
  \usepackage{mathrsfs}
\usepackage{mathcomp}

  \long\def\comment#1{ }

  \newcommand{\beq}{\begin{eqnarray}}
  \newcommand{\eeq}{\end{eqnarray}}
  
 \def\simge{\mathrel{%
   \rlap{\raise 0.511ex \hbox{$>$}}{\lower 0.511ex \hbox{$\sim$}}}}
\def\simle{\mathrel{
   \rlap{\raise 0.511ex \hbox{$<$}}{\lower 0.511ex \hbox{$\sim$}}}}

\vspace{1.5cm}

\title{\rm \LARGE \bf Fermionic Schwinger--Keldysh Propagators from AdS/CFT}

\author{G. C. Giecold \\Institut de Physique Th\'eorique,
CEA Saclay, CNRS (URA 2306),
 F-91191 Gif-sur-Yvette, France\\
  E-mail: \email{gregory.giecold@cea.fr}}

\abstract{The Herzog and Son prescription \cite{Herzog:2002pc} for computing real--time Green functions for finite temperature gauge theories from their gravity dual is generalized to fermions. These notes explain how such an extension involves properties of spinors in a curved, complexified space--time.}

\begin{document}

\section{Introduction}

The gauge--string duality relates some classes of field theories to dual string theories in specified background space--times \cite{D'Hoker:2002aw}.
While string theory in a curved background does not generally lends itself to tractable calculations and even to our understanding, its low--energy supergravity limit is far more compliant. In the case of asymptotically AdS spaces, the dual field theories are in the large 't Hooft coupling limit.
The AdS/CFT correspondence thus provides a framework for understanding strongly--coupled gauge theories. Recent work has been devoted to computing transport coefficients \cite{CasalderreySolana:2006rq, Gubser:2006nz, CasalderreySolana:2007qw, Son:2009vu, Giecold:2009cg, deBoer:2008gu} and gaining insight into dynamic and nonequilibrium settings \cite{Chesler:2008hg, Giecold:2009wi} from the correspondence. Most of them required a real--time formulation of finite temperature field theory. The way real--time correlators can be derived in AdS/CFT is hinted by the following analogy. There is a doubling of the degrees of freedom in the Schwinger--Keldysh real--time prescription (reviewed in Section 2 of this paper). On the other hand, the Penrose diagrams of asymptotically AdS spacetimes with a black hole\footnote{whose temperature is that of the dual gauge theory, according to the AdS/CFT correspondence} exhibit two boundaries (cf. Figure 1. below), on which the dual gauge theory fields live. This conjecture was proved by Herzog and Son \cite{Herzog:2002pc}. They showed how the $2 \times 2$ matrix of two--point correlation functions for a scalar field and its doubler partner field is reproduced from the AdS dual supergravity action. Their work also made it clear that the thermal nature of black hole physics gives rise to the thermal nature of its dual field theory. In these notes, we would like to extend their work and find out how black hole physics gives rise to real--time correlators of fermionic operators in a dual finite--temperature field theory. While deriving real--time propagators of vector field operators from AdS/CFT is an obvious extension of the scalar field case expounded in \cite{Herzog:2002pc}, the case of fermions proves less straightforward. The analysis presented below relies on a treatment of spinor fields in curved space--times and on their transformation laws under global symmetry transformations.
We begin by reviewing in the next section the Schwinger--Keldysh formalism for real--time finite temperature field theory. This section collects results usually dispersed among the literature, as this formalism is usually exposed for a scalar field. We present the $2 \times 2$ matrix of two--point correlation functions for a fermionic operator and its link with the retarded and advanced propagators. Section 3 reviews how the retarded Green function for a fermionic operator at strong coupling can be computed from the dual supergravity spinor classical action in AdS/CFT. Section 4 is devoted to a short review of spinors in curved and complexified space--times. These are important for the analysis of Section 5 where we review the relationship between positive-- and negative--energy modes of a wave equation and analyticity conditions in the complex Kruskal planes of the underlying background. These conditions lead to the real--time propagators for fermionic operators from the dual boundary action in the gauge--gravity duality.
 
\begin{figure}
\centerline{
\includegraphics[angle=270,width=0.6\textwidth]{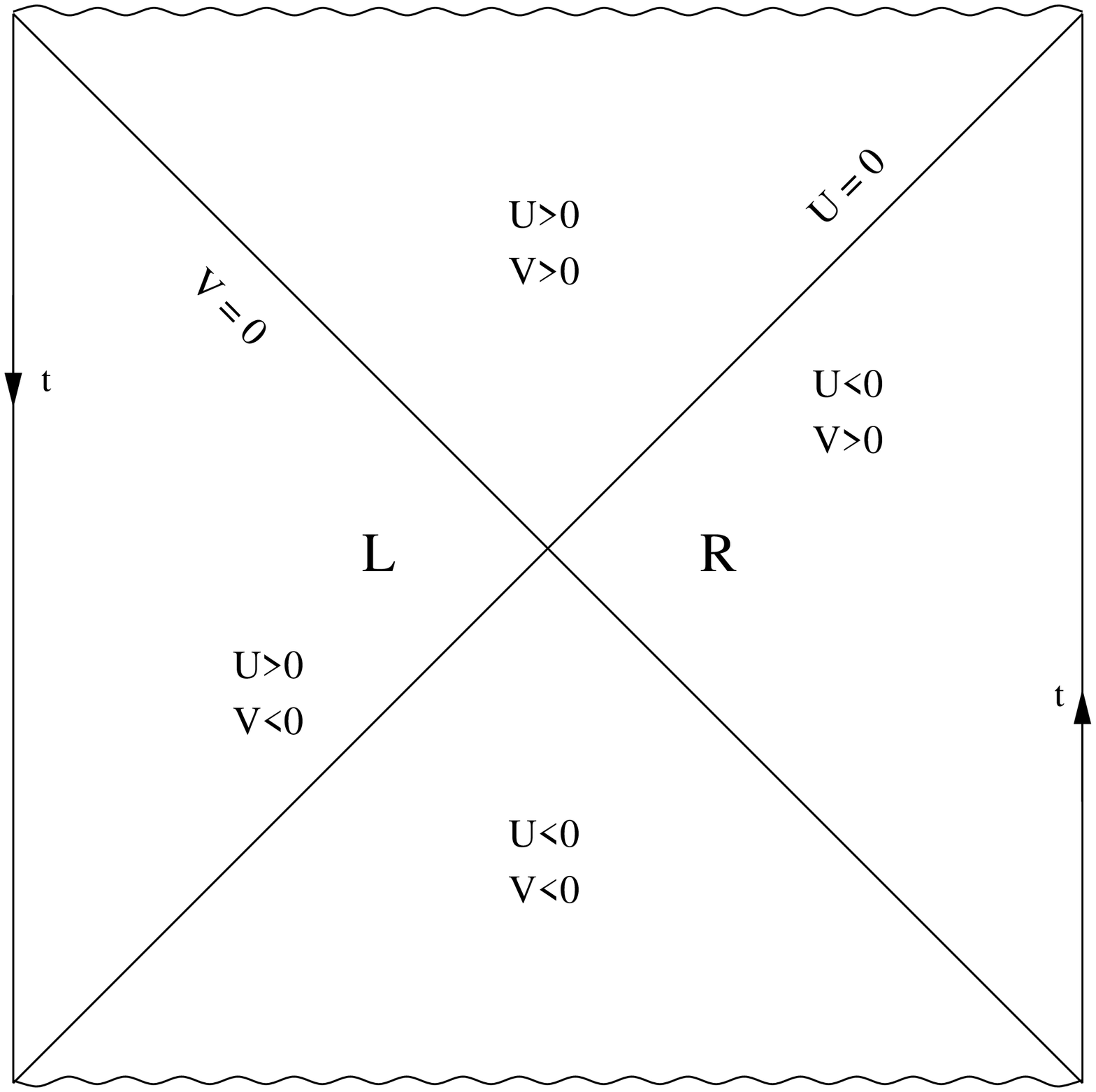}}
\caption{\sl Kruskal diagram for AdS--Schwarzschild black hole}
\end{figure}

Recently, there has been a sustained interest in fermions from theories with gravity duals. 
In \cite{Akhavan:2009ns} the two--point function for a fermionic operator in a non--relativistic conformal field theory is computed. The gravity dual corresponds to fermions propagating in a background with the Schr\"odinger isometry. 
The authors of \cite{Cubrovic:2009ye} argue that the gauge--gravity correspondence proves a useful tool for exploring fermionic quantum phase transitions. The retarded fermion Green function is found from an analysis of the solutions to the Dirac equation and its quasi--normal modes in an AdS Reissner--Nordstr\"om black hole. Real--time correlators for non--relativistic holography have been considered recently in \cite{Leigh:2009eb}, where the construction of  \cite{Skenderis:2008dh, Skenderis:2008dg} is involved. For an explanation of how their construction generalizes the one due to \cite{Herzog:2002pc} used here to the case of distinct sources on the R and L boundaries of a Penrose diagram, see \cite{vanRees:2009rw}.
It would be interesting to apply the approach contained in the present notes to more general geometries, possibly duals to non--relativistic conformal field theories \cite{Son:2008ye, Herzog:2008wg, Maldacena:2008wh, Adams:2008wt}. 
This paper offers to explain how Schwinger--Keldysh $n$--point functions can be computed from string theory, thus emphasizing the latter as a relevant approach to tackle some models or phases of condensed matter physics. 

\section{Review of Schwinger--Keldysh formalism for fermions}

The Schwinger--Keldysh prescription allows for a study of real--time Green functions by introducing a contour $\mathcal{C}$ in the complex time plane \cite{LeBellac:1996, Landsman:1986uw}. Fields live on this time contour. The forward and return contour of the path are labelled by indices $i_1$ and $i_2$ respectively. The idea is that the quantum dynamics does the doubling of the degrees of freedom required for describing non--equilibrium states. The starting point $I$ at time $t_i$ and the ending point $B$ at $t_i - i \beta$ are identified and fermionic fields are such that $\Upsilon_I = - \Upsilon_B$. In the remainder of this paper, the conventions for the propagators are those of \cite{KadanoffBaym:1989}.

\setlength{\unitlength}{1cm}
\begin{picture}(16,6)
\put(3,5){\line(1,0){8}}
\put(6.76,5){\vector(1,0){.3}}
\put(6.9,5.2){$\mathcal{C}$}
\put(3,5){\circle*{.2}}
\put(2.5,4.9){A}
\put(3.3,5.2){$t_{i}$}
\put(11.1,5.2){$t_{f}$}
\put(11,5){\line(0,-1){1}}
\put(11.1, 3.8){$t_{f} - i \sigma$}
\put(11,4){\line(-1,0){8}}
\put(7.24,4){\vector(-1,0){.3}}
\put(3,4){\line(0,-1){1.5}}
\put(3,2.5){\circle*{.2}}
\put(3,3.5){\vector(0,-1){.3}}
\put(3.3,2.7){$t_{i} - i \beta$}
\put(2.5,2.4){B}
\put(3.3,1){$\text{Figure 2: The Schwinger--Keldysh contour}$}
\end{picture}

The action splits into contributions from the four parts of the contour:
\begin{align}\label{Action splitting}
S &= \int_{\mathcal{C}} dt_{\mathcal{C}} L(t_{\mathcal{C}}), \nonumber \\
   &= \int_{t_i}^{t_f} dt L(t) - i \int_0^{\sigma} d\tau L(t_f - i \tau) - \int_{t_i}^{t_f} dt L(t - i \sigma) - i \int_{\sigma}^{\beta} d\tau L(t_i - i \tau),
\end{align}
where 
\beq\label{Action psi psi bar}
L(t) = \int d^{d-1}\vec{x} \mathcal{L}\left[ \Upsilon(t,\mathbf{x}), \bar{\Upsilon}(t, \mathbf{x}) \right].
\eeq
The generating functional is defined as
\begin{align}\label{Generating functional}
Z &= \int \mathcal{D}\Upsilon \mathcal{D}\bar{\Upsilon} \exp \left( i S + i \int_{t_i}^{t_f} d^d x \bar{\eta}_1 \Upsilon_1 + i \int_{t_i}^{t_f} d^d x \bar{\Upsilon}_1 \eta_1 - i \int_{t_i}^{t_f} d^d x \bar{\eta}_2 \Upsilon_2 - i \int_{t_i}^{t_f} d^d x \bar{\Upsilon}_2 \eta_2 \right), \nonumber \\
&
\end{align}
where the sources $\eta_{1,2}$ and the fields are such that
\beq\label{Source field} \left\{
\begin{array}{lll}
\eta_1 (t,\mathbf{x}) = \eta (t, \mathbf{x}), & \ \ \ \Upsilon_1(t,\mathbf{x}) = \Upsilon(t,\mathbf{x}), \\
\eta_2 (t,\mathbf{x}) = \eta (t- i\sigma, \mathbf{x}), & \ \ \ \Upsilon_2 (t,\mathbf{x}) = \Upsilon(t-i\sigma,\mathbf{x}).\\
\end{array} \right.
\eeq
The same relations hold for their conjugates.
The contour--ordered Green functions are mapped into a matrix whose components are indexed by the position on the contour:
\beq\label{Green matrix}
i G(j,k) = \frac{1}{i^2} \frac{\delta^2 \ln Z\left[ \eta_{1,2}, \bar{\eta}_{1,2} \right] }{\delta \eta_j \delta \eta^{\dagger}_k} = i 
\begin{pmatrix}
G_{11} & G_{12} \\
G_{21} & G_{22} \\
\end{pmatrix}.
\eeq
The time in the components of this matrix of Green function is standard time and in the operator formalism
\beq\label{Green un deux} \left\{
\begin{array}{lll}
G_{11}(t,\mathbf{x}) = - i \langle T \Upsilon(t,\mathbf{x}) \Upsilon^{\dagger}(0) \rangle, & \ \ \ G_{12} (t,\mathbf{x}) = + i \langle \Upsilon^{\dagger}(0) \Upsilon (t,\mathbf{x}) \rangle, \\
G_{21} (t,\mathbf{x}) = - i \langle \Upsilon (t,\mathbf{x}) \Upsilon^{\dagger}(0) \rangle, & \ \ \ G_{22}(t,\mathbf{x}) = - i \langle \hat{T} \Upsilon(t,\mathbf{x}) \Upsilon^{\dagger}(0) \rangle.
\end{array} \right.
\eeq
Note the sign reversal in $G_{12}$ as compared to the case where the fields are bosonic. $T$ and $\hat{T}$ denote the time--ordering and anti--time--ordering operators. The fields are taken in the Heisenberg picture.
Those Schwinger--Keldysh correlators are related to the retarded and advanced Green functions through
\beq\label{G ret advanced}
\left\{
\begin{array}{ll}
G_R (x - y) = - i \theta(x^0 - y^0) \langle \left\{ \Upsilon(x), \Upsilon^{\dagger}(y) \right\} \rangle, \\
G_A (x - y) = + i \theta(y^0 - x^0) \langle \left\{ \Upsilon(x), \Upsilon^{\dagger}(y)  \right\} \rangle .    
\end{array} \right.
\eeq
The relation
\beq\label{G R conjug G A}
G_R(x-y) = G^*_A(y-x)
\eeq
is valid, irrespective of the fields obeying a Bose--Einstein or Fermi--Dirac statistics.
Using \eqref{Green un deux}, \eqref{G ret advanced} and the completeness relation for a complete set of state results in
\beq\label{Green un deux ret adv} \left\{
\begin{array}{lll}
G_{11}(k) = - \text{Re} G_R(k) + i \tanh(\frac{k^0}{2T}) \text{Im} G_R(k), \\
G_{12}(k) = -\frac{2i e^{\sigma k^0}}{1+ e^{\beta k^0}} \text{Im} G_R(k), \\
G_{21}(k) = \frac{2i e^{(\beta - \sigma)k^0}}{1 + e^{\beta k^0}} \text{Im} G_R(k), \\
G_{22}(k) = \text{Re} G_R(k) + i \tanh(\frac{k^0}{2T}) \text{Im} G_R(k).
\end{array} \right.
\eeq
When $\sigma = \frac{\beta}{2}$ --- a value which will naturally appear in the following --- $G_{12}(k) = - G_{21}(k)$. $\sigma = 0$ yields $G_{21}(k) = G^{>}(k)$ and $G_{12}(k) = G^{<}(k)$. Since $G_{21}(k) - G_{12}(k) \mid_{\sigma = 0} = 2i \text{Im} G_R(k)$, the relationship 
\beq\label{G R A sup inf}
G_R(k) - G_A(k) = G^{>}(k) - G^{<}(k)
\eeq   
holds as required whatever the quantum statistics of the fields under consideration.
The main purpose of these notes is to show how the above relations \eqref{Green un deux ret adv} are derived in AdS/CFT. Herzog and Son \cite{Herzog:2002pc} obtain analogous relations for a scalar field. As it turns out, when one tries to extend their result to correlators of fermionic operators one runs one runs into some specific and non completely obvious issues. 

\section{Review of fermionic retarded correlators in AdS/CFT}

This section reviews fermions in AdS/CFT \cite{Henningson:1998cd, Henneaux:1998ch}. The prescription for computing retarded fermionic Green functions draws on the approach of Iqbal and Liu \cite{Iqbal:2009fd, Iqbal:2008by}, where conjugate momenta for supergravity fields are defined with respect to a $r$--foliation. This is suggestive of some sort of stochastic quantization interpretation of the AdS/CFT correspondence.\\
Consider a boundary fermionic operator $\mathcal{O}$ whose gravity dual is a  spinor field $\Psi$. The bulk background space--time metric
\beq\label{Background metric}
ds^2 = g_{rr} dr^2 + g_{\mu \nu} dx^{\mu} dx^{\nu}
\eeq 
is subjected to the asymptotically--AdS conditions 
\beq\label{AdS asymptotic}
g_{tt}, g_{ii} \sim r^2, \ \ \ g_{rr} \sim 1/r^2, \ \ \ r \rightarrow \infty.
\eeq
The $d$--dimensional boundary is a $r \rightarrow \infty$.
The AdS/CFT prescription for computing $n$--point functions of a quantum field theory from a classical supergravity action goes as
\beq\label{Witten prescription}
\left\langle \exp\left[ \int d^{d}x \left( \bar{\chi}_0 \mathcal{O} + \bar{\mathcal{O}} \chi_0 \right) \right] \right\rangle_{QFT} = e^{i S_{SUGRA} [\chi_0, \bar{\chi}_0]},
\eeq
where $\chi_0 = \lim_{r \rightarrow \infty} r^{d - \Delta} \Psi$ and $\Delta$ is the scaling dimension of $\mathcal{O}$, related to the mass $m$ of the bulk spinor. Imposing such a Dirichlet boundary condition on spinors requires care, especially given that this condition must relate a spinor in the bulk in $d+1$ dimensions to one in $d$ space-time dimensions on the boundary.
Given that our focus is on Green functions, the quadratic part of the action for $\psi$ meets our purpose. It is given by
\beq\label{S quadra}
S = i \int_M d^{d+1}x \sqrt{-g} \left( \bar{\Psi} \Gamma^{M} D_{M} \Psi - m \bar{\Psi} \Psi \right) + S_{\partial M},
\eeq
with $\bar{\Psi} = \Psi^{\dagger} \Gamma^{t}$.
The covariant derivative is specified by the spin connection $\omega_{abM}$:
\beq\label{Covariant deriv}
D_{M} = \partial_{M} + \frac{1}{4} \omega_{abM} \Gamma^{ab},
\eeq
where $\Gamma^{ab} = \Gamma^{[ a} \Gamma^{b ]}$. Upper--case letters stand for abstract space--time indices, while lower--case ones denote tangent frame indices. The two are linked through a choice of vielbein $e^{a}_{M}$, defined by $G_{MN} = e^{a}_M e^{b}_N \eta_{ab}$, with $\eta_{ab}$ a $d+1$--dimensional Minkowski metric of signature $(-, +, +, ..., +)$ such that 
\beq\label{Gamma anticomm}
\left\{ \Gamma^{a}, \Gamma^{b} \right\} = 2 \eta^{ab}.
\eeq 
The inverse vielbein satisfy $\eta_{ab} = G_{MN} e_a^M e_b^N$. The spin connection components are given by $\omega_{abc} = e_a^M \omega_{bcM}$ and $\omega_{abc} = \frac{1}{2}\left( \mathfrak{C}_{bca} + \mathfrak{C}_{acb} - \mathfrak{C}_{abc} \right)$, where $\left[ e_a, e_b \right] = \nabla_{e_a}e_b - \nabla_{e_b}e_a = \mathfrak{C}_{ab}^c e_c$. Alternatively $\omega_{abM}$ can be viewed as the components of 1--forms $\mathbf{\omega}^M$ in Cartan's structure equations \cite{MST:gravitation}. Whenever a specific index appears as a label on a Gamma matrix, it refers to that particular index in the tangent frame. 
For $d$ even, a convenient choice for the bulk Gamma matrices is
\beq\label{Gamma d even}
\Gamma^{\mu} = \gamma^{\mu}, \ \ \ \Gamma^{r} = \gamma^{d+1},
\eeq
$\gamma^{\mu}$ being the boundary gamma matrices and $\gamma^{d+1}$ being proportional to their product.
When $d$ is odd, it is appropriate to choose
\beq\label{Gamma d odd}
\Gamma^{\mu} = 
\begin{pmatrix}
0 & \gamma^{\mu} \\
\gamma^{\mu} & 0 \\
\end{pmatrix},
\ \ \ \Gamma^{r} = 
\begin{pmatrix}
1 & 0 \\
0 & -1\\
\end{pmatrix}.
\eeq
They satisfy the Clifford algebra.
So, for general $d$
\beq\label{Psi plus minus}
\Psi = \Psi_+ + \Psi_-, \ \ \ \Psi_{\pm} = \Gamma_{\pm} \Psi, \ \ \ \Gamma_{\pm} = \frac{1}{2}\left( 1 \pm \Gamma^{r} \right),
\eeq
with $\Psi_{\pm}$ being opposite chirality Weyl spinors when $d$ is even, and $d$--dimensional Dirac spinors for $d$ odd. Whatever the value of $d$ the number of components of the fermionic operator $\mathcal{O}$ is always half that of its dual $\Psi$.
Quite generally \cite{Polchinski:book2}, Appendix B, $D$--dimensional gamma matrices are constructed by noticing that for $D$ even, increasing $D$ by $2$ doubles the size of the Dirac matrices. They can therefore be constructed iteratively, starting in $D=2$ with
\beq\label{Dirac 2 d}
\Gamma^0 = \begin{pmatrix} 0 & 1 \\ -1 & 0 \end{pmatrix}, \ \ \ \Gamma^1 = \begin{pmatrix} 0 & 1 \\ 1 & 0 \end{pmatrix},
\eeq 
to the following in $D = 2 k + 2$
\beq\label{Dirac pair}
\Gamma^{\mu} = \gamma^{\mu} \bigotimes \begin{pmatrix} -1 & 0 \\ 0 & 1 \end{pmatrix}, \ \ \Gamma^{D-2} = I \bigotimes \begin{pmatrix} 0 & 1 \\ 1 & 0 \end{pmatrix}, \ \ \Gamma^{D-1} = I \bigotimes \begin{pmatrix} 0 & -i \\ i & 0 \end{pmatrix}.
\eeq
Here, $\gamma^{\mu}$, $\mu = 0, ..., D-3$ are $2^k \times 2^k$ gamma matrices and I is the $2^k \times 2^k$ identity matrix.
When $D$ is odd, simply add $\Gamma^D = \Gamma$ or $-\Gamma$ to the set of $D-1$ gamma matrices. Note that from our conventions for the metric and anti--commutation relations the $0$--component of gamma matrices is anti--hermitian while other matrices are hermitian.
In order to solve the equations of motion near the boundary and find the scaling dimension $\Delta$ of the fermionic operator $\mathcal{O}$, one refers to the usual Frobenius procedure of trying for solutions of the type $r^{-\rho} \sum_{n = 0}^{\infty} \Psi_n(t,x^{i}) / r^{n}$. $\Psi_n$ are boundary spinors. 
Consider for instance the case of pure AdS, $ds^2 = r^2 (-dt^2 + d\mathbf{x}^2) + \frac{dr^2}{r^2}$. Setting $e^{\mu} = r dx^{\mu}$, $e^r = \frac{dr}{r}$, the non--vanishing spin coefficients 
\beq\label{spin coeff pure AdS}
\omega_{tr} = -r dt, \ \ \ \omega_{ir} = r dx^{i}.
\eeq
The Dirac equation
\beq\label{Dirac equation}
\left[ \slashed{D} - m \right] \Psi = 0, 
\eeq
with $\slashed{D} = \Gamma^M D_M$,
becomes
\beq\label{Dirac pure AdS}
r \Gamma^{r} \partial_{r} \Psi + \frac{i}{r} \Gamma \ . \ k \Psi + \frac{d}{2} \Gamma^{r} \Psi - m \Psi = 0, 
\eeq
where $\Gamma \ . \ k = \gamma^{\mu} k_{\mu}$.
Then $\rho$ must be set to $\rho = \frac{d}{2} \pm m$ and $\Psi_0$ is annihilated by $\frac{1}{2} \left( 1 \pm \Gamma^{r} \right)$, respectively. Incidentally, the scaling dimension is therefore found to be $\Delta = \frac{d}{2} + m$. The leading asymptotic behaviour of $\Psi_{\pm}$ is then 
\beq\label{Psi asymptotic}
\Psi_{+}(k, r) = \chi_0 (k) r^{m - \frac{d}{2}} + \lambda_0 (k) r^{-\frac{d}{2} - m - 1}, \Psi_{-}(k, r) = \psi_0 (k) r^{-\frac{d}{2} - m} + \mu_{0}(k) 
r^{m - \frac{d}{2} - 1}
\eeq
Note that the dominant term (when $m \geq 0$) has been denoted $\chi_0$ on purpose, as a reference to the source for the dual operator $\mathcal{O}$ in \eqref{Witten prescription}: 
\beq\label{Source asymptotic}
\lim_{r \rightarrow \infty} r^{d - \Delta} \Psi_+ = \chi_0.
\eeq
Inserting this back into their equation of motion yields
\begin{align}\label{A B C D relations}
\left\{
\begin{array}{ll}
\psi_0 (k) = - \frac{i \gamma \ . \ k}{k^2} (1 + 2m) \lambda_0 (k) ; \\
\mu_0 (k) = - \frac{i \gamma \ . \ k}{1 - 2m} \chi_0 (k). \\
\end{array} \right.
\end{align}
If one then demands that the solution be regular in the whole of AdS space, it turns out that $\chi_0$ and $\psi_0$ are not independent. Note that this is not apparent from the analysis presented above where $\chi_0$ and $\psi_0$ are the boundary values of the fields $\Psi_+$ and $\Psi_-$. A general solution to the Dirac equation near the boundary is a superposition of those fields. However the Dirac equation can be solved exactly in a few cases, including pure AdS. In this latter case, suitable solutions are given by
\begin{align}\label{Dirac solution pure AdS}
\Psi_+ = \left\{
\begin{array}{lll}
r^{-\frac{d+1}{2}} K_{m + \frac{1}{2}} \left( \frac{\sqrt{\mathbf{k}^2 - \omega^2 }}{r} \right) \kappa_+, \ \ \ k^2 > 0, \\
r^{-\frac{d+1}{2}} H^{(1)}_{m + \frac{1}{2}} \left( \frac{\sqrt{\omega^2 - \mathbf{k}^2}}{r} \right) \kappa_+, \ \ \ \omega > \sqrt{\mathbf{k}^2}, \\
r^{-\frac{d+1}{2}} H^{2}_{m + \frac{1}{2}} \left( \frac{\sqrt{\omega^2 - \mathbf{k}^2}}{r} \right) \kappa_+, \ \ \ \omega < - \sqrt{\mathbf{k}^2},\\
\end{array} \right.
\end{align}
$\kappa_+$ denoting a constant spinor.
Regularity as $r \rightarrow 0$ then imposes that
\beq\label{psi chi relation}
\psi_0 (k) = - \frac{i \gamma \ . \ k}{k^2} \frac{(\frac{k}{2})^{2m} \Gamma(\frac{1}{2} - m)}{\Gamma(\frac{1}{2} + m)} \chi_0 (k).
\eeq
More generally $\psi_0$ and $\chi_0$ are related by a matrix $\mathcal{S}$:
\beq\label{S matrix}
\psi_0 (k) = \mathcal{S} (k) \chi_0 (k).
\eeq
Regularity in the bulk and a given boundary condition $\chi_0$ for $\Psi_+$ when $m \geq 0$ then uniquely determine a solution $\Psi$ to the classical equations of motion. Similar relations apply for $\bar{\Psi}_{+}$ and $\bar{\Psi}_{-}$. Note that relations such as \eqref{psi chi relation} or \eqref{S matrix} are on--shell relations. Other off--shell histories contributing to the variational principle can be constructed as superpositions of independent $\Psi_+$ and $\Psi_-$. We now turn to a discussion of the variational principle. The boundary term in \eqref{S quadra} will be determined from stationarity of the action. That one cannot fix all the components of $\Psi$ and $\bar{\Psi}$ but must rather set conditions on, say, $\chi_0$, $\bar{\chi}_0$, and leave $\psi_0$, $\bar{\psi}_0$ free to vary\footnote{On--shell they become functions of the boundary conditions $\chi_0$, $\bar{\chi}_0$.}, stems from the Dirac equation being first order in derivatives. Varying with respect to $\Psi$, $\bar{\Psi}$ the Euclidean action
\beq\label{Dirac action}
S = - \int_M d^{d+1} x \sqrt{g} \bar{\Psi} \left( \frac{1}{2} \left( \stackrel{\rightarrow}{\slashed{D}} - \stackrel{\leftarrow}{\slashed{D}} \right) - m \right) \Psi,
\eeq
one finds a surface term, i.e. $\delta S = C_{\partial M} +$ bulk term, where the bulk term involves radial derivatives and is proportional to the equations of motions, while
\begin{align}\label{C surface term}
C_{\partial M} =& \frac{1}{2} \int_{\partial M} d^d x \left( \delta \bar{\psi}_0 \chi_0 + \bar{\chi}_0 \delta \psi_0 \right)(x), \nonumber \\
                       =&  \delta \left\{ \frac{1}{2} \int_{\partial M} d^d x \left( \bar{\psi}_0 \chi_0 + \bar{\chi}_0 \psi_0 \right)(x) \right\},
\end{align}
given that $\delta \chi_0 = 0$ and $\delta \bar{\chi}_0 = 0$, as explained above.
Since $C_{\partial M}$ does not vanish, that the action must be stationary requires that one adds to it a boundary term $S_{\partial M}$ such that $C_{\partial M} = - \delta S_{\partial M}$. Actually, since they are fixed, one may add any function of $\chi_0$ and $\bar{\chi}_0$ to the action without breaking the stationarity condition. However conditions such as locality, absence of derivatives and invariance under the asymptotically AdS symmetries seem to uniquely select a boundary term \cite{Henneaux:1998ch}. Hence, after another Wick--rotation to go back to a Lorentzian signature,
\beq\label{Boundary surface term}
S_{\partial M} = -i \int_{\partial M} d^d x \sqrt{gg^{rr}} \bar{\Psi}_{+} \Psi_{-}.
\eeq
The factor of $g^{rr}$ entering the square root comes from the vielbein.
We now review the prescription derived by Iqbal and Son \cite{Iqbal:2009fd, Iqbal:2008by} for computing retarded propagators. It amounts to taking Euclidean canonical momenta conjugate to $\Psi_{\pm}$ with respect to a $r$--foliation:
\beq\label{Pi plus minus momenta}
\Pi_+ = - \sqrt{gg^{rr}} \bar{\Psi}_-, \ \ \ \Pi_- = - \sqrt{gg^{rr}} \bar{\Psi}_+.
\eeq 
Then \eqref{Psi asymptotic} and \eqref{Source asymptotic} result in
\begin{align}\label{1P fctn}
\langle \mathcal{O} \rangle_{\chi_0} &= - \lim_{r \rightarrow \infty} r^{\Delta - d} \Pi_+, \nonumber \\
                                                          &= \bar{\psi}_0.
\end{align}
From \eqref{S matrix}, i.e. $\psi_0 = \mathcal{S} \chi_0$, and analytic continuation, one obtains retarded correlators in Lorentzian signature:
\beq\label{Correlators Lorentzian}
G_R(k) = i \mathcal{S}(k) \gamma^{t}.
\eeq
The $\gamma^{t}$ gamma matrix arises since $G_R \sim \langle \mathcal{O} \mathcal{O}^{\dagger} \rangle$, rather than $\langle \mathcal{O} \bar{\mathcal{O}} \rangle$. 

\section{Spinors in complexified space--time}

In order to generalize the work of Herzog and Son \cite{Herzog:2002pc} to fermions, it is necessary to consider spinors in curved space--time. Besides, a potential difficulty arises given that \cite{Herzog:2002pc} relies crucially on analycity of complexified Kruskal coordinates. \cite{Penrose:book2}, section 6.9, has a few pages devoted to spinors in complex space--times. \cite{Penrose:book1, Newman:1961qr} provide a complementary treatment of spinors and twistors. A spin space $\aleph$ of complex dimension two comes with each point of the underlying space--time manifold. The members of such a space are negative--chirality, dotted, say, Weyl spinors. Undotted spinors are members of the complex conjugate space $\bar{\aleph}$. The manifold being complexified, $\aleph$ and $\bar{\aleph}$ must be viewed as independent spaces, so that pairs of spinors $\xi^{\alpha}$ and $\bar{\xi}^{\dot{\alpha}}$ which previously determined one another under complex conjugation, are replaced by a pair of independent such spinors. A complexified space--time originates from a real underlying space--time by allowing its coordinates to take on complex values and by extending the metric coefficients analytically to the complex domain. Note that this is distinct from a complex space--time where generally no subspace can be singled out as real. Defining a spinor basis or dyad $ \left\{ \varsigma^{\alpha},\imath^{\alpha} \right\}$ for $\aleph$ and $\left\{ \bar{\varsigma}^{\dot{\alpha}}, \bar{\imath}^{\dot{\alpha}} \right\} $ for $\bar{\aleph}$, each comes bestowed with its own indices--lowering $\epsilon_{\alpha \beta}$ spinor or $\bar{\epsilon}_{\dot{\alpha}\dot{\beta}}$ spinor, such that $\epsilon ( \varsigma, \imath ) = \gamma$ and $\bar{\epsilon} ( \bar{\varsigma}, \bar{\imath} ) = \bar{\gamma}$. When $\gamma = 1$, the dyad is called a spin--frame. Associated with any spin frame is a null tetrad 
\beq\label{T basis}
l = \varsigma \bar{\varsigma}, \ \ \ m = \varsigma \bar{\imath}, \ \ \ n = \imath \bar{\imath}, \ \ \ \bar{m} = \imath \bar{\varsigma}, 
\eeq 
which spans the tensor product space $\aleph \bigotimes \bar{\aleph}$. This illustrates the standard connection between world--tensor indices $a$ as a pair of spinor indices, one dotted and one undotted.
From $\epsilon$ and $\bar{\epsilon}$, a symmetric metric on $T$ is built such that
\beq\label{complex metric}
g(l,n) = 1, \ \ \ g(m,\bar{m}) = -1,
\eeq
with all other scalar products vanishing, i.e.
\beq\label{complex metric indices}
g_{ab} = \epsilon_{\alpha \beta} \epsilon_{\dot{\alpha}\dot{\beta}}.
\eeq
$\aleph \bigotimes \bar{\aleph}$ endowed with this scalar product has the structure of a tangent space to a complexified manifold. 
In the Newman--Penrose formalism a basis for the tangent space is a null tetrad consisting of two real vectors and one complex--conjugate pair of vectors. Consider the complexified manifold obtained from a metric of type \eqref{Background metric} describing a geometry with a horizon. Let $U$ and $V$ label its Kruskal coordinates. In the context of an asymptotically AdS black hole geometry they are introduced below around \eqref{Kruskal coordinates explicit}. Staying general for now, a basis for the tangent space is given by four null vectors
\beq\label{M tangent space basis}
\frac{\partial}{\partial U}, \ \ \ \frac{\partial}{\partial V}, \ \ \ \frac{\partial}{\partial \zeta} = e^{i \phi} \cot \frac{\theta}{2}, \ \ \ \frac{\partial}{\partial \bar{\zeta}} = e^{- i \phi} \cot \frac{\theta}{2},
\eeq
with $\frac{\partial}{\partial \zeta}$ parameterizing the anti--celestial sphere. It is related to $\frac{\partial}{\partial \bar{\zeta}}$ by an antipodal map.
Let us map $l$, $n$, $m$ and $\bar{m}$ to $\frac{\partial}{\partial U}$, $\frac{\partial}{\partial V}$, $\frac{\partial}{\partial \zeta}$ and $\frac{\partial}{\partial \bar{\zeta}}$, respectively. The spinors $\varsigma$ and $\imath$ are then associated with the null vectors $\frac{\partial}{\partial U}$ and $\frac{\partial}{\partial V}$, respectively. Given that the vector $U \frac{\partial}{\partial U}$ is parallely transported along the orbits of $\frac{\partial}{\partial U}$ and those of $\frac{\partial}{\partial V}$ (all the same for $V \frac{\partial}{\partial V}$)
\beq\label{Parallel transport}
\nabla_{\frac{\partial}{\partial U}} \left[ U \frac{\partial}{\partial U} \right] = 0, \ \ \ \nabla_{\frac{\partial}{\partial V}} \left[ U \frac{\partial}{\partial U} \right] = 0, 
\eeq
the spinors $\sqrt{V}\imath$ and $\sqrt{-U} \varsigma$ or $\sqrt{-V} \imath$ and $\sqrt{U} \varsigma$ are parallely transported across the full U and V complex planes.
Equation \eqref{Parallel transport} results from $U$ = const. and $V$ = const. each corresponding to null geodesics. Geodesics are defined such that their tangent vector field $n$ obeys $\nabla_n n = 0$. Generally any vector $T$ is said to be parallely transported along the orbits of vector field $n$ if $\nabla_n T$ points in the direction of $T$. This happens above with $n = \frac{\partial}{\partial U}$ and $T = U \frac{\partial}{\partial U}$:
$\nabla_{\frac{\partial}{\partial U}} \left[ U \frac{\partial}{\partial U} \right] = U \nabla_{\frac{\partial}{\partial U}} \frac{\partial}{\partial U} + \left( \frac{\partial}{\partial U}\left[ U \right] \right) \frac{\partial}{\partial U}$. The parametrization of a geodesic can be rescaled so that $\alpha = 0$, though, $\alpha$ being the proportionality factor between $\nabla_n T$ and $T$ above.\\
This choice of spinor basis will appear naturally in Section 5. $\sqrt{-U}$ and the likes are pivotal in generating Fermi--Dirac distribution functions. We should also note that even though the AdS setting presented below involves spinors in a bulk geometry of dimension five, the present discussion on spinors in a complexified four--dimensional space--time is of relevance due to the decomposition \eqref{Psi plus minus} of a general bulk spinors evolving in $d+1$ dimensions into $d$--dimensional spinors. The latter are used in the following as in, e.g., \cite{Henningson:1998cd, Henneaux:1998ch, Iqbal:2009fd}, and while in five dimensions an extra fifth basis vector should appear in \eqref{M tangent space basis}, it is irrelevant for the present purpose. $\Psi_{\pm}$ spinors will be expanded in the basis constructed out of Kruskal coordinates and $\imath, \varsigma$. 

\section{Real--time correlators from gravity}

While the results about to be derived below should be applicable to a broader class of finite--temperature field theories with a gravity dual, we focus on theories arising from non--extremal D3 branes. Boundary values of supergravity fields in the resulting $\text{AdS}_5$ background provide the $N \rightarrow \infty$, $g^2_{YM} N \rightarrow \infty$ limit of $\mathcal{N} = 4$ SU(N) supersymmetric Yang--Mills correlators. 
In the near--horizon limit the metric on a stack of non--extremal D3 branes reads\footnote{Setting $R =1$ for convenience.}
\beq\label{Metric r Schwarz}
ds^2 =  \left( \pi T R r \right)^2 \left( -f(r) dt^2 + d\mathbf{x}^2 \right) + R^2 \frac{dr^2}{r^2 f(r)},
\eeq
where $f(r) = 1-\frac{1}{r^4}$. The boundary is a $r \rightarrow \infty$ and the horizon at $r = 1$. Here, $T$ is the Hawking temperature of this AdS--Schwarzschild black hole and R is the radius of AdS. The analysis of \cite{Herzog:2002pc} relies on the behaviour of a scalar field in this background. 
Near the horizon, $\frac{r}{\pi T} = 1 + \epsilon$, solutions to the wave equation behave as $e^{ik^0r^*}$ and its conjugate, with $k^0 = \omega$ and $r^*$ being the tortoise coordinate:
\beq\label{Tortoise equation}
\frac{dr^*}{dr} = \frac{1}{\pi T} \frac{1}{r^2 f(r)}, \ \ \ r^* = \frac{1}{2 \pi T} \left( \arctan(r) + \log \sqrt{\frac{r-1}{r+1}} \right).
\eeq
The Kruskal coordinates are defined as
\beq\label{Kruskal coordinates explicit}
\left\{
\begin{array}{ll}
U = - \frac{e^{- 2 \pi T (t + r^*)}}{2 \pi T}, \\
V = \frac{e^{2 \pi T (t - r^*)}}{2 \pi T}.
\end{array} \right.
\eeq
The Penrose diagram of Figure 1. is constructed from these coordinates. 
The retarded and advanced solutions comport themselves as
\beq\label{Solution asymptotic}
\left\{
\begin{array}{ll}
e^{-i \omega t} f(k, r) \sim e^{- \frac{i \omega}{2 \pi T} \ln(V)}, \ \  \text{in--falling} , \\
e^{-i \omega t} f^{*}(k, r) \sim e^{\frac{i \omega}{2 \pi T} \ln(-U)},  \ \ \text{out--going}.
\end{array} \right.
\eeq
When we considered solutions to the wave equation, we were working in the R--quadrant, $U < 0$, $V > 0$. However, as explained in \cite{Herzog:2002pc}
if one extends the mode functions to the complex $U$ and $V$ planes, one finds that positive--frequency solutions to the wave equation are analytic in the lower $U$ and $V$ complex planes. A solution is composed of only negative--frequency modes provided it is analytic in the upper $U$ and $V$ planes.
With regard to the modes of \eqref{Solution asymptotic} one then requires that the solution be analytic in the lower $V$ plane and the upper $U$ plane.
Since in the $r-t$ coordinates one can solve the wave equation independently in the R and L regions of the Penrose diagram one obtains the following set of mode functions in each quadrants:
\beq\label{U modes R L quadrants}
\begin{matrix}
u_{R,i}(k) = \left\{
\begin{array}{ll}
e^{-i \omega t} f(k, r), \text{   in R} \\
0, \text{   in L}
\end{array} \right. & u_{L,i}(k) = \left\{
\begin{array}{ll}
0, \text{   in R} \\
e^{-i \omega t} f(k, r), \text{   in L}
\end{array} \right. \\
u_{R,o}(k) = \left\{
\begin{array}{ll}
e^{-i \omega t} f^{*}(k, r), \text{   in R} \\
0, \text{   in L} 
\end{array} \right. & u_{L,o}(k) = \left\{
\begin{array}{ll}
0, \text{   in R} \\
e^{-i \omega t} f^{*}(k, r), \text{   in L}
\end{array} \right.
\end{matrix}
\eeq
Only two linear combinations can be built which meet the above criterium on holomorphicity. These are
\beq\label{U out and in combi}
\left\{
\begin{array}{ll}
u_o = u_{R,o} + \alpha_o u_{L,o} , \\
u_i = u_{R,i} + \alpha_i u_{L,i}.
\end{array} \right.
\eeq
From the behaviour close to the horizon of the solutions and the analyticity requirement, the in--going and out--going cross--connecting functions $\alpha_i$ and $\alpha_o$ are constrained to be
\beq\label{Alpha in, qui est out}
\left\{
\begin{array}{ll}
\alpha_o = e^{\frac{\pi \omega}{2}}, \\
\alpha_i = e^{-\frac{\pi \omega}{2}}.
\end{array} \right.
\eeq
In order to carry a similar analysis to the case of fermions, one must first check that solutions to the Dirac equation in an AdS--Schwarzschild background \eqref{Metric r Schwarz} behave as $e^{-i\omega r*}$ and its conjugate.
In such a background\footnote{Setting $R =1$, $\pi T = 1$ for convenience.} the spin coefficients are 
\beq\label{Spin coeff Schwarz}
\omega_{tr} = - r \left(1 + \frac{1}{r^4} \right) dt, \ \ \ \omega_{ir} = r \sqrt{f} dx^i.
\eeq
The Dirac equation then reads
\beq\label{Dirac eq expanded} \left\{
\begin{array}{ll}
\left[ - \frac{i \omega}{\sqrt{f}} \gamma^t + i \vec{\gamma} \ . \ \vec{k} \right] A(m) \Psi_+(k,r) = \left[ - \frac{\omega^2}{f} + \mathbf{k}^2 \right] \Psi_-(k,r), \\
\left[ - \frac{i \omega}{\sqrt{f}} \gamma^t + i \vec{\gamma} \ . \ \vec{k} \right] A(-m) \Psi_-(k,r) = - \left[ - \frac{\omega^2}{f} + \mathbf{k}^2 \right] \Psi_+(k,r),
\end{array} \right.
\eeq
where $A(m) = r \left[ r \sqrt{f(r)} \partial_r + \frac{d-1}{2} \sqrt{f(r)} + \frac{(1 + \frac{(\pi T)^4}{r^4})}{2 \sqrt{f(r)}} - m \right]$. Focusing on the terms relevant for the near--horizon behaviour, solutions of the type $\frac{r}{\sqrt[4]{f(r)}} e^{\pm i \omega r^*}$ satisfy these equations. Note that it is crucial that the $\Gamma^0$ matrix be anti--hermitian. This leading near--horizon behaviour of solutions to the Dirac equation in a curved background is reminiscent of the forms of the solutions found in \cite{Unruh:1974bw} in the course of this study of second--quantization for neutrino fields in a Kerr background.
From the previous discussion on parallely--transported spinors in the complex U and V planes, one is led to consider the following set of mode functions in each quadrant:
\beq\label{Psi modes R L quadrants}
\begin{matrix}
\psi_{R,i} = \left\{
\begin{array}{ll}
e^{-i \omega t} \sqrt{V} f(k, r) \imath, \text{   in R} \\
0, \text{   in L}
\end{array} \right. & \psi_{L,i} = \left\{
\begin{array}{ll}
0, \text{   in R} \\
e^{-i \omega t} \sqrt{-V} f(k, r) \imath, \text{   in L}
\end{array} \right. \\
\psi_{R,o} = \left\{
\begin{array}{ll}
e^{-i \omega t} \sqrt{-U} f^{*}(k, r) \varsigma, \text{   in R} \\
0, \text{   in L} 
\end{array} \right. & \psi_{L,o} = \left\{
\begin{array}{ll}
0, \text{   in R} \\
e^{-i \omega t} \sqrt{U} f^{*}(k, r) \varsigma, \text{   in L}
\end{array} \right.
\end{matrix}
\eeq
The mergers $f(k,r) \begin{Bmatrix} \sqrt{-U} \\ \sqrt{V} \end{Bmatrix}$ behave as $\frac{r}{\sqrt[4]{f(r)}} e^{\pi T t} e^{i\omega r^*}$, as required for solutions to the Dirac equation, except for the extra $e^{\pm \pi T t}$ term, which could be inserted in the definitions of the modes in \eqref{Psi mode expansion} below.
As for the scalar case reviewed above, the conditions that positive--frequency solutions are analytic in the lower $U$ and $V$ complex planes and negative--energy modes are analytic in their upper counterparts leads to the following linear combinations
\beq\label{Psi out and in combi}
\left\{
\begin{array}{ll}
\psi_o = \psi_{R,o} + \beta_o \psi_{L,o}, \\
\psi_i = \psi_{R,i} + \beta_i \psi_{L,i}.
\end{array} \right.
\eeq
The behaviour of the solutions close to the horizon fixes
\beq\label{Beta in, out}
\left\{
\begin{array}{ll}
\beta_o = i e^{\frac{\pi \omega}{2}}, \\
\beta_i = - i e^{- \frac{\pi \omega}{2}}.
\end{array} \right.
\eeq
The out--going and in--going solutions in \eqref{Psi out and in combi} form a basis for a spinor field defined over the full Kruskal plane of the AdS--Schwarzschild geometry
\beq\label{Psi mode expansion}
\Psi_{-}(r) = \sum_k \left[ a(\omega, \mathbf{k}) \psi_o (k, r) + b(\omega, \mathbf{k}) \psi_i (k, r) \right]. 
\eeq
In accordance with our discussion on the variational principle for spinor fields in AdS/CFT we do not expand the $\Psi_+$ field. Its leading--order part in an expansion near the boundary is fixed. One must fix the ``position'' and leave the ``momentum'' free to vary in a set of canonically conjugate pairs given by $\bar{\chi}_0$ and $\psi_0$.
The coefficients $a(\omega, \mathbf{k})$, $b(\omega, \mathbf{k})$ are determined by requiring that \eqref{Psi mode expansion} approaches $\Psi^R_-(k)$ and $\Psi^L_-(k)$ on their respective boundaries:
\beq\label{a et b vs psi R L}
\left\{
\begin{array}{ll}
\left[ a(\omega, \mathbf{k}) \sqrt{-U} \varsigma + b(\omega, \mathbf{k}) \sqrt{V} \imath \right]_{r_{\partial M}} = \Psi^R_-(k) ,\\
- \left[ a(\omega, \mathbf{k}) e^{\pi \omega} \sqrt{U} (-) \varsigma - b(\omega, \mathbf{k}) \sqrt{-V} (-)\imath \right]_{r_{\partial M}} = -i e^{-\frac{\pi \omega}{2}} \Psi^L_-(k),
\end{array} \right.
\eeq
The function $f(k,r)$ is normalized such that $f(k, r_{\partial M}) = 1$ at the boundary. The overall minus sign on the r.h.s. of the second equation results from the effect on spinor fields of time reversal from going to the R--quadrant to the L one\footnote{Our conventions for the Clifford algebra differ from those of \cite{Srednicki:qft}. This affects in particular the $\Gamma^5$ matrix. Hence the overall sign flip in \eqref{Spinor time reversal} as compared to the more familiar equation (40.32) from \cite{Srednicki:qft}.}:
\beq\label{Spinor time reversal}
T^{-1}\Psi_{+,\alpha}(x)T = - \Psi_{+,\alpha}(\mathcal{T}x), \ \ \ T^{-1}\Psi^{\dot{\alpha}}_{-} T = + \Psi_{-,\dot{\alpha}}(\mathcal{T}x).
\eeq
Also, recall that raising or lowering a spinor index comes with a minus sign. The meaning of \eqref{a et b vs psi R L} is as a set of two equations for two unknown spinors, $a(\omega, \mathbf{k}) \sqrt{-U} \varsigma$ and $b(\omega, \mathbf{k}) \sqrt{V} \imath$. The second equation in \eqref{a et b vs psi R L} involves the same unknown spinors but in the L quadrant, which introduces extra $\sqrt{-1}$. Taking care of those additional factors of $i$ which occur in going from $\begin{Bmatrix} \sqrt{U} \\ \sqrt{- V} \end{Bmatrix}$ to $\begin{Bmatrix} \sqrt{- U} \\ \sqrt{V} \end{Bmatrix}$, \eqref{a et b vs psi R L} leads to 
\beq\label{a et b individuels}
\left\{
\begin{array}{ll}
a(\omega, \mathbf{k}) \sqrt{-U} \mid_{r_{\partial M}} \varsigma = \frac{1}{e^{\pi \omega}+1} \left[ \Psi^R_- (k)+ e^{\frac{\pi \omega}{2}} \Psi^L_-(k) \right], \\
b(\omega, \mathbf{k}) \sqrt{V} \mid_{r_{\partial M}} \imath = \frac{1}{e^{\pi \omega}+1} \left[ e^{\pi \omega} \Psi^R_-(k) - e^{\frac{\pi\omega}{2}} \Psi^L_-(k) \right].
\end{array} \right.
\eeq
Here, $n(\omega) = \frac{1}{e^{\beta \omega} + 1}$ is the Fermi--Dirac distribution.
A computation of real--time Green functions from the standard AdS/CFT prescription is now in order. The classical boundary action in $r-t$ coordinates is
\begin{align}\label{S bndry R L}
S_{\partial M} &= - i \int_{\partial M} \frac{d^d k}{(2 \pi)^d} \sqrt{-gg^{rr}} \bar{\Psi}_+(-k,r) \Psi_-(k,r) \nonumber \\
&= - i \int \frac{d^d k}{(2 \pi)^d} \sqrt{-gg^{rr}} \bar{\Psi}_+(-k,r) \Psi_-(k,r) \mid_{r_R} - i \int \frac{d^d k}{(2 \pi)^4} \sqrt{-gg^{rr}} \bar{\Psi}_+(-k,r) \Psi_-(k,r) \mid_{r_L} 
\end{align}
For a scalar field the second integral would have come with the opposite sign. Here, however one must recall that spinor fields behave non--trivially when they cross the L quadrant where time ordering is reversed. While $\bar{\Psi}\Psi$ is invariant under time--reversal, what we are really considering instead is an expression where $\bar{\Psi}_+$ is fixed whereas $\Psi_-$ is free to vary. The unusual sign is associated with the latter's transformation under time reversal, \eqref{Spinor time reversal}. 
Using \eqref{Psi mode expansion} and \eqref{a et b individuels} the boundary action becomes
\begin{align}\label{S bndry expanded}
S_{\partial M} =& - i \int \frac{d^4 k}{(2 \pi)^4} \sqrt{-gg^{rr}} n(\omega) u_{R,o}(k) \left[ \bar{\Psi}^R_+(-k) \Psi^R_-(k) + e^{\frac{\beta \pi}{2}} \bar{\Psi}^R_+(-k) \Psi^L_-(k) \right] \nonumber \\
& - i \int \frac{d^4 k}{(2 \pi)^4} \sqrt{-gg^{rr}} n(\omega) u_{R,i}(k) \left[ e^{\frac{\beta \omega}{2}} \bar{\Psi}^R_+(-k) \Psi^R_-(k) - e^{\beta \omega} \bar{\Psi}^R_+(-k) \Psi^L_-(k) \right] \nonumber \\
& + i \int \frac{d^4 k}{(2 \pi)^4} \sqrt{-gg^{rr}} n(\omega) u_{L,o}(k) e^{\frac{\beta \omega}{2}} \left[ \bar{\Psi}^L_+(-k) \Psi^R_-(k) + e^{\frac{\beta \omega}{2}} \bar{\Psi}^L_+(-k) \Psi^L_-(k) \right] \nonumber \\
& - i \int \frac{d^4 k}{(2 \pi)^4} \sqrt{-gg^{rr}} n(\omega) u_{L,i}(k) e^{-\frac{\beta \omega}{2}} \left[ e^{\beta \omega} \bar{\Psi}^L_+(-k) \Psi^R_-(k) - e^{\frac{\beta \omega}{2}} \bar{\Psi}^L_+(-k) \Psi^L_-(k) \right].
\end{align}
Equations \eqref{Correlators Lorentzian} and \eqref{G R conjug G A}, which for fermionic operators reads $G_{A,ab}(k) = G^*_{R,ba} = - i S^*(k) \gamma^t_{ba} = i S^*(k) \gamma^t_{ab}$ -- from $\Gamma^0$ being anti--hermitian, cf., e.g., \eqref{Dirac pair} -- and the near--boundary expansions of $\Psi^{R,L}_{\pm}$ yield
\beq\label{Real Green gravity} \left\{
\begin{array}{llll}
G_{RR}(k) = - i \left[ n(\omega) e^{\beta \omega} (-i) G_R(k) + n(\omega) (-i) G_A(k) \right] = - \text{Re} G_R(k) + i \tanh(\frac{\omega}{2T}) \text{Im} G_R(k), \\
G_{LL}(k) = - i \left[ - n(\omega) (-i) G_R(k) - n(\omega) e^{\beta \omega} (-i) G_A(k) \right] = \text{Re} G_R(k) + i \tanh(\frac{\omega}{2T}) \text{Im} G_R(k), \\
G_{RL}(k) = - i n(\omega) e^{\frac{\beta \omega}{2}} \left[ (-i) G_R(k) - (-i) G_A(k) \right] =  -\frac{2i e^{\frac{\beta \omega}{2}}}{1+ e^{\beta \omega}} \text{Im} G_R(k), \\
G_{LR}(k) = - i n(\omega) e^{\frac{\beta \omega}{2}} \left[ - (-i) G_R(k) + (-i) G_A(k) \right] = \frac{2i e^{\frac{\beta \sigma}{2}}}{1+ e^{\beta \omega}} \text{Im} G_R(k).
\end{array} \right.
\eeq
These are the Schwinger--Keldysh propagators \eqref{Green un deux ret adv} with $\sigma = \frac{\beta}{2}$ for a fermionic operator dual to the supergravity field $\Psi$. One can redefine $\Psi^L(k)$ to obtain real--time propagators with arbitrary $0 < \sigma < \beta$.
Let us illustrate how \eqref{Real Green gravity} is obtained and focus on $G_{RR}(k)$.\\
The relevant terms from the boundary action are\\
$-i n(\omega) \sqrt{-gg^{rr}} \left[ u_{R,i}(k) e^{\beta \omega} \bar{\Psi}^R_+(-k)\Psi^R_-(k) + u_{R,o}(k) \bar{\Psi}^R_+(-k)\Psi^R_-(k) \right]$.\\
The first one comes with an in--going mode. One then uses the near--boundary expansion for $\Psi^R_{\pm}$ and the relation \eqref{S matrix}, i.e. $\psi_0 (k) = \mathcal{S} (k) \chi_0 (k)$, to translate this term into an expression proportional to the retarded propagator. On the other hand, the second term in brackets involves an out--going mode. It must be associated with an advanced Green function \cite{Son:2002sd}. One then writes $\sqrt{-gg^{rr}} \bar{\Psi}^R_+(-k)\Psi^R_-(k) = \sqrt{-gg^{rr}} \bar{\Psi}^R_-(k)\Psi^R_+(-k)$ which is equal to $\mathcal{S}^*(k) \bar{\chi}^R_0(k)\chi^R_0(-k) = \mathcal{S}^*(k) \bar{\chi}^R_0(-k)\chi^R_0(k)$. This finally leads to the stated result. 
We have thus checked that Schwinger--Keldysh correlators for fermionic operators and the standard relations among them and the retarded, advanced and symmetric propagators hold in AdS/CFT by taking functional derivatives of boundary part of the dual supergravity action. This prescription can be generalized to higher--point functions of a fermionic operator, provided the non--quadratic parts of the action for its dual supergravity spinor field are known.

\subsection*{Acknowledgments}
This work was supported in part by Contrat de Formation par la Recherche, CEA Saclay.
Thanks are due to Tae--Joon Cho, from DAMTP, for references on spinors and twistors.

\end{document}